\documentclass[]{aastex6}

\usepackage{multirow}
\fullcollaborationName{The Friends of AASTeX Collaboration}

\begin{document}
 \fontsize{10pt}{10pt}{Comments: Accepted for publication in The Astrophysical Journal. 9 pages, 9 figures, 4 tables.}
\title{Fermi-LAT Detection of GeV $\gamma$-Ray Emission from the highly asymmetric shell supernova remnant: SNR G317.3-0.2}

\author{Yunchuan Xiang\altaffilmark{1}, Zejun jiang\altaffilmark{1} and Yunyong Tang\altaffilmark{2}}

\altaffiltext{1}{Department of Astronomy, Yunnan University, and Key Laboratory of Astroparticle Physics of Yunnan Province, Kunming, 650091, China, xiang{\_}yunchuan@yeah.net, zjjiang@ynu.edu.cn} 

\altaffiltext{2}{School of Physical Science and Technology, Kunming University, Kunming 650214, China, tangyunyong888@163.com}


\begin{abstract}
In this paper, we report the first extended GeV $\gamma$-ray emission, at a significant level of $\sim$ 8.13$\sigma$, from the region of the supernova remnant (SNR) SNR G317.3-0.2 by analyzing $\sim$ 12.2 years of Fermi Large Area Telescope (Fermi-LAT) Pass 8 data in the work. The best-fit position of the new $\gamma$-ray source matches that of the 843 MHz radio energy band of SNR G317.3-0.2, and there is no significant variability of the photon flux of the corresponding light curve (LC) in the data for the 12.2 year period; therefore, by excluding other known $\gamma$-ray sources or candidates within a 2$\sigma$ error radius from the best-fit position of SNR G317.3-0.2, we suggest that the $\gamma$-ray source is likely to be a GeV counterpart of SNR G317.3-0.2. 
\end{abstract}

\section{Introduction} \label{sec:intro}
\citet{Whiteoak1996} reported features at 843 MHz for SNR G317.3-0.2, as a shell supernova remnant (SNR), based on observations performed using the Molonglo Observatory Synthesis Telescope (MOST). SNR G317.3-0.2 consists of two asymmetric and opposing arcs, with a flux of 5.2 Jy. The 843 MHz flux density map showed a strong asymmetry between the SE lobe and  NW lobe, and the flux density
of the SE lobe was brighter than that of the NW lobe. Their findings suggested that the diffuse emission of SNR G317.3-0.2 was likely to be associated with
the nearby ionized region from H \uppercase\expandafter{\romannumeral2} GAL 316.8-0.1. Its surface luminosity was $5.8 \times 10^{-21} \rm W\ m^{-2}\ Hz^{-1}\ s^{-1}$ at a distance of 9.7 kpc at 1 GHz \citep{Case1998}. \citet{Stupar2011} found an excellent match between the H$\alpha$  and 4850 MHz radio emissions in the central and southwest region. The evidence suggests that the SNR could
be an optical counterpart.

\citet{Acero2016} analyzed the GeV $\gamma$-ray emission of SNR G317.3-0.2 using Fermi-LAT data from August 4, 2008, to August 4, 2011.  However, they did not find a significant $\gamma$-ray emission. The 95\% and 99\% confidence upper limits with two power-law spectral indices ($\Gamma =$ 2.0 or 2.5) from the region were provided in their work. Thus far,  X-ray and TeV energy bands have not been observed precisely for SNR G317.3-0.2.

\citet{West2016} used a physically motivated model, which involves the evolution of the morphology of supernova remnants in the magnetic field, to simulate the magnetic field distribution of SNR G317.3-0.2 according to its 843 MHz radio morphology. They found a strong asymmetry, and their results resembled the distribution of the 843 MHz flux density. They considered that the asymmetry was likely to be caused by the vector of the magnetic field along the X-axis in a three-dimensional coordinate system. In the magnetic field region where SNRs enter, it can produce radio synchrotron emission with higher intensity, resulting in the bright limbs of SNRs \citep{van1962,Whiteoak1968}.


With the accumulation of photons, more precise diffuse $\gamma$-ray backgrounds, and an improved dataset, we
revisit the $\gamma$-ray radiation of SNR G317.3-0.2 by analyzing approximately 12.2 years of the Fermi-LAT Pass 8 data \citep{4FGL}. We find the likely GeV $\gamma$-ray radiations from SNR G317.3-0.2. The 843 MHz flux density map with a
higher spatial resolution of 43$^{''}$ of SNR G317.3-0.2 presents a clear and asymmetric structure, which enabled us to
better study the spatial distribution of its GeV $\gamma$-ray radiation by using the Fermi-LAT analysis method provided by
the Fermi Science Support Center\footnote{http://fermi.gsfc.nasa.gov/ssc/data/analysis}.  In addition, the highly asymmetric structure of SNR G317.3-0.2 at 843 MHz,
which makes it be an excellent source for studying the asymmetric evolution of SNR and the unknown origin of cosmic
rays in the Milky Way in the future \citep{Aharonian2004,
 Helder2012,Ackermann2013,Bao2018,Ferrand2019,Orlando2020}.

In this paper, the data analysis procedures are introduced in Section 2, and the analysis results are presented in Section 3.  The detection and likely origin of GeV radiation are discussed and conclusions are provided in Section 4.

\section{Data Reduction} \label{sec:data-reduction}
We used Fermi Science Tools version {\tt v11r5p3}\footnote{http://fermi.gsfc.nasa.gov/ssc/data/analysis/software/} to perform the analysis.
The Pass 8 ``Source'' event class (evtype = 3 and evclass = 128) and  the instrumental response function (IRF) ``P8R3{\_}SOURCE{\_}V2'' were used to analyze the $\gamma$-ray emission from the source.
Events with zenith angles above $90^{\circ}$ were excluded to reduce the contamination of the Earth Limb. 
We set the energy range of data recorded from August 4, 2008, to October 24, 2020 (mission elapsed time 239557427-625268550) as 2 GeV to 500 GeV to avoid a large point spread function (PSF) and the pollution of the galactic diffuse background in the lower energy band. 
Region of interest (ROI) was a $20^{\circ}\times 20^{\circ}$ and was centered at the position R.A. = 222$^{\circ}$.43, Decl. =  -59$^{\circ}$.77 from SIMBAD\footnote{http://simbad.u-strasbg.fr/simbad/}.
Assuming a point source with a power-law spectral model at the location of SNR G317.3-0.2, we calculated its best-fit position to be R.A., decl. = 222.51, -59.75 with a 68\% (95\%) error circle of 0$^{\circ}$.046 (0$^{\circ}$.074) using {\tt gtfindsrc}, and we replaced the original position of SNR G317.3-0.2 with the best-fit position. 
Then, we marked the new $\gamma$-ray source as SrcX for all subsequent analyses.
The script {\tt make4FGLxml.py}\footnote{https://fermi.gsfc.nasa.gov/ssc/data/analysis/user/} was used to generate the source model file containing all sources within an ROI of 30$^{\circ}$ around SNR G317.3-0.2 from the Fermi Large Telescope Fourth Source Catalog \citep[4FGL;][]{4FGL}.
  Furthermore, a point source with a power-law spectral model in the best-fit position of SrcX was added to the model file. 
The tool {\tt gtlike}\footnote{https://fermi.gsfc.nasa.gov/ssc/data/analysis/scitools/references.html} was used to fit the data by following the data analysis method of the Fermi Science Support Center\footnote{http://fermi.gsfc.nasa.gov/ssc/data/analysis/scitools/}.
All spectral indexes and normalizations from sources within $5^{\circ}$ were set as free. Moreover, the normalizations from the Galactic diffuse emission ({\tt gll{\_}iem{\_}v07.fits}) and the isotropic extragalactic emission ({\tt iso{\_}P8R3{\_}SOURCE{\_}V2{\_}v1.txt})\footnote{http://fermi.gsfc.nasa.gov/ssc/data/access/lat/BackgroundModels.html} were also set as free.

\section{Results} \label{sec:data}
\begin{figure}[!h]
  \includegraphics[width=60mm,height=60mm]{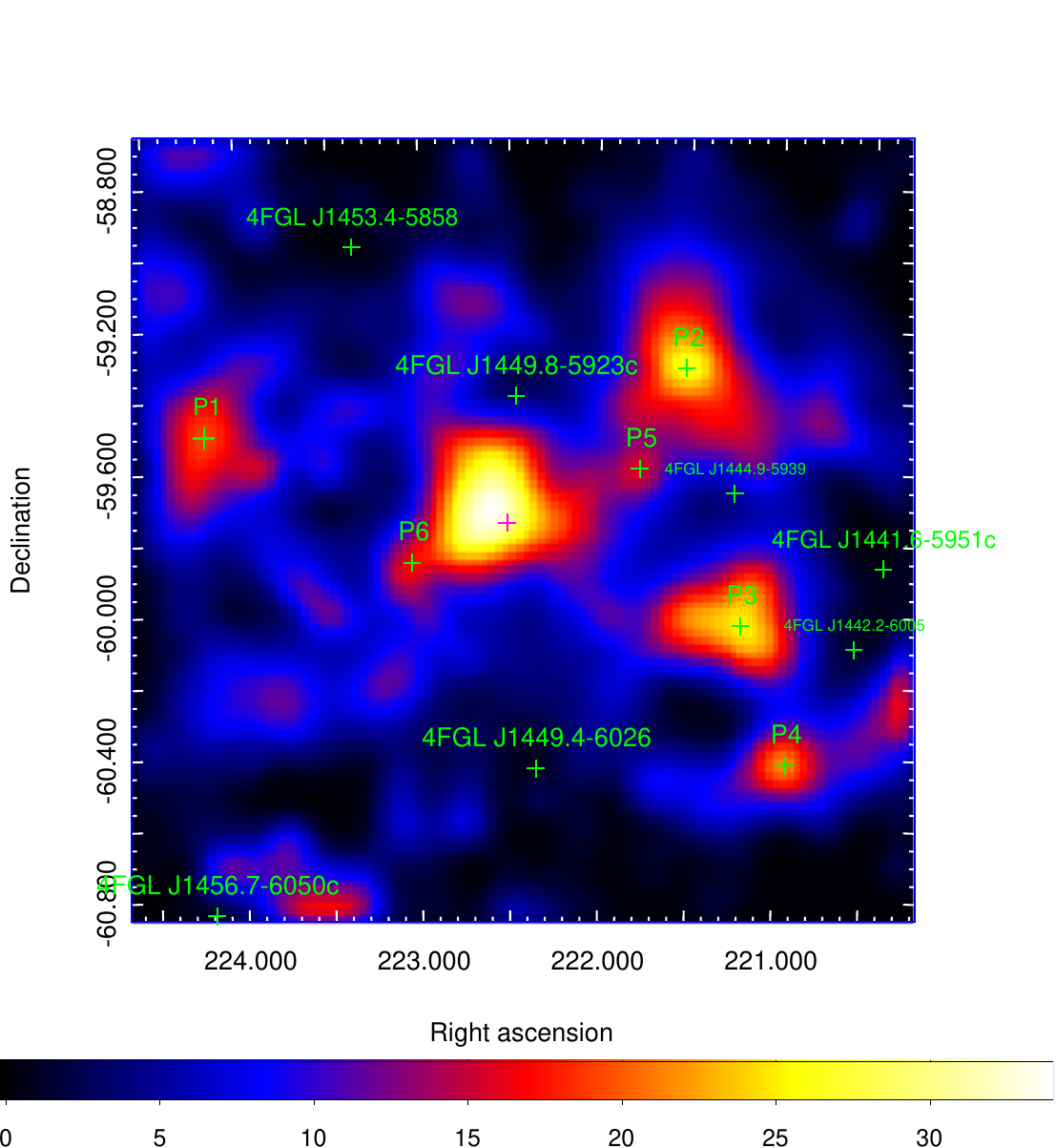}
  \includegraphics[width=60mm,height=60mm]{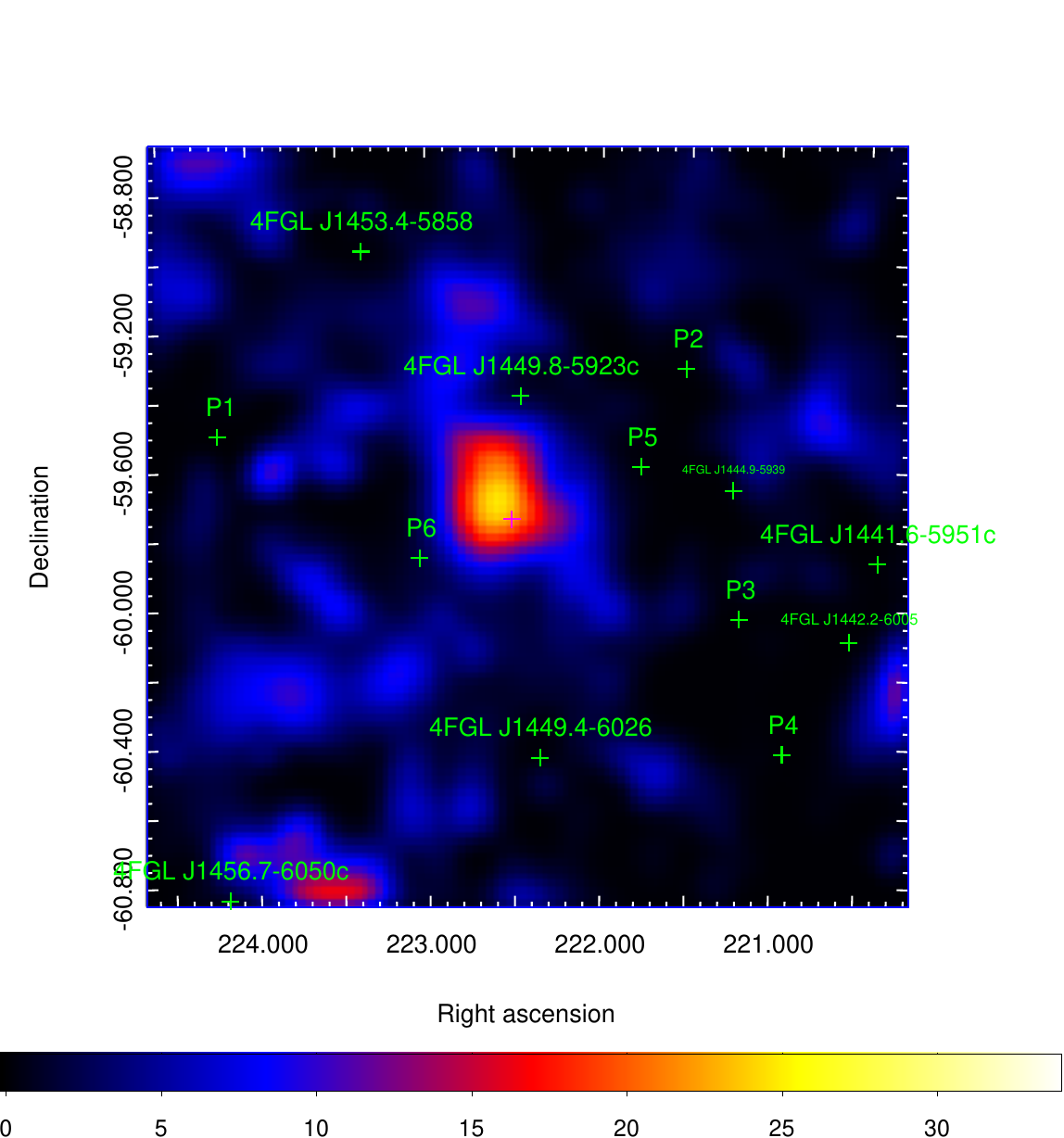}
  \includegraphics[width=60mm,height=60mm]{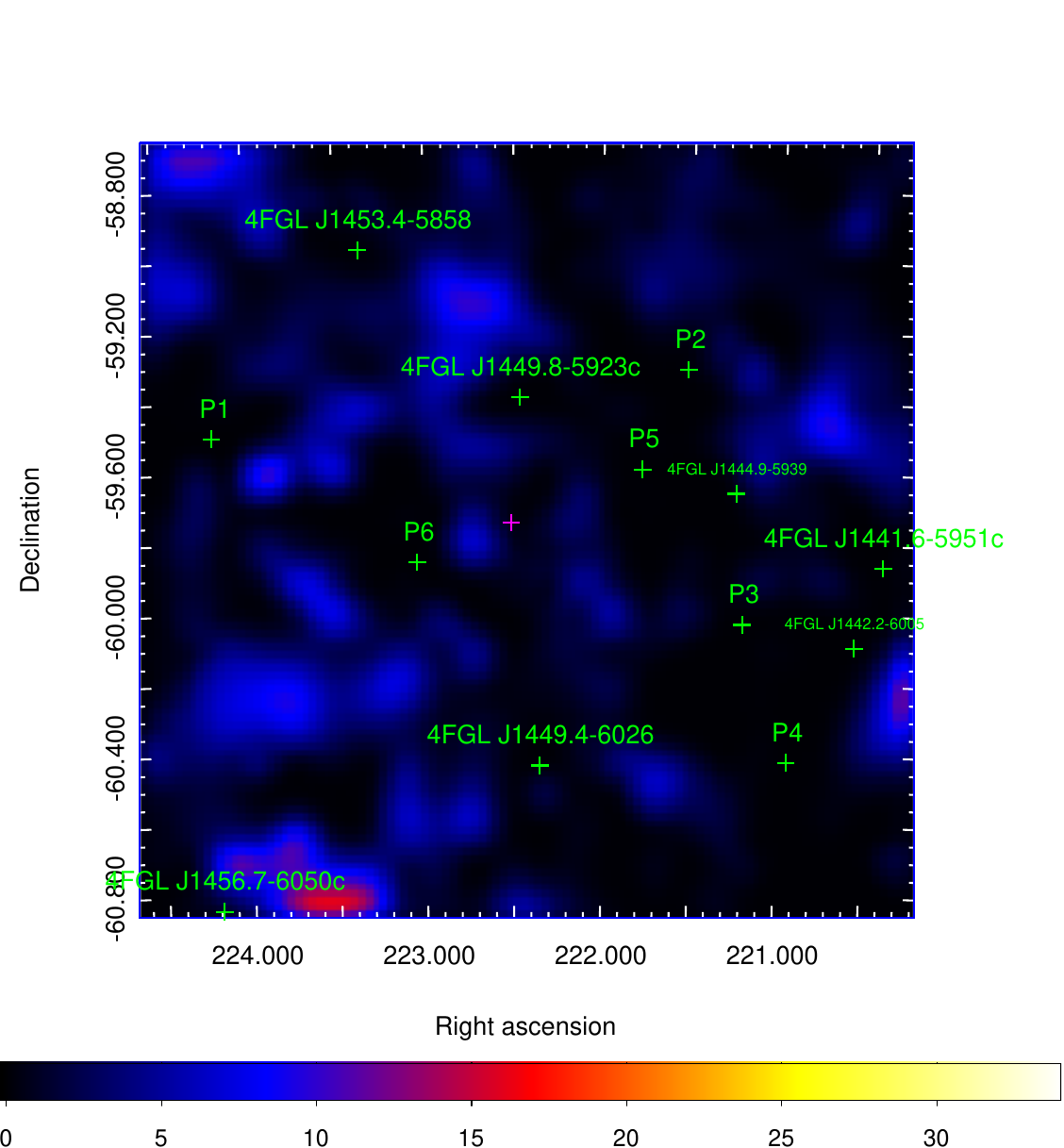}
 \caption{
Three TS maps with 0$^{\circ}$.02 pixel size in the 2-500 GeV energy band center at the position of SrcX marked as a magenta cross in a $2^{\circ}.2\times2^{\circ}.2$ region. 
Left panel: TS map including all sources from 4FGL.
Middle panel: TS map after subtracting P1-P6.  Right panel: TS map after subtracting all sources containing SrcX and P1-P6.
 A Gaussian function with a kernel radius of $0^{\circ}.3$ was used to smooth these maps.}
    \label{Fig1}
\end{figure}

\begin{figure}[!h]
\centering
 \includegraphics[width=\textwidth, angle=0,width=100mm,height=100mm]{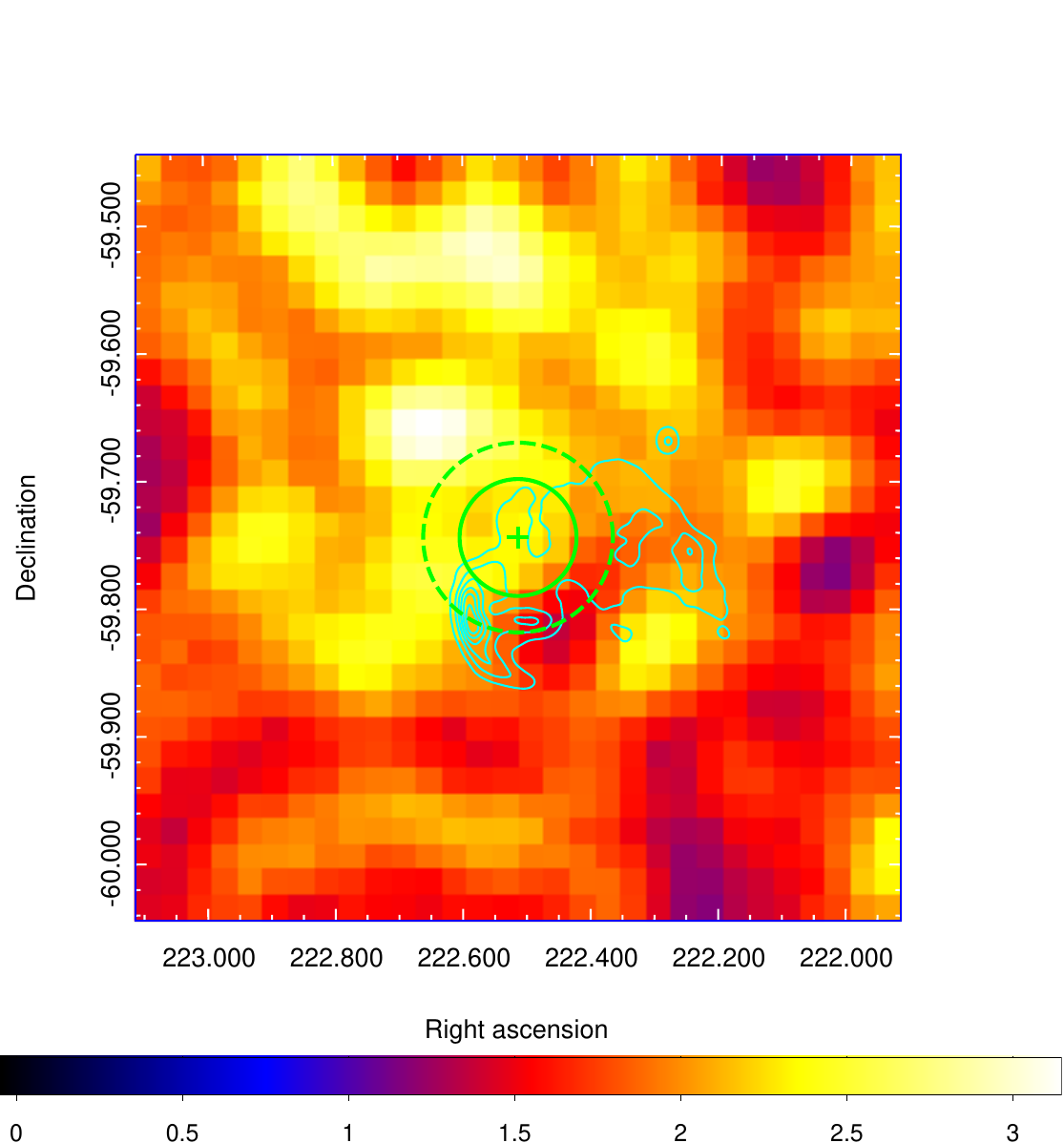} %
 \caption{Counts map with $0^{\circ}.02$ pixel size in the 2-500 GeV energy band centered at the position of SrcX marked as a green
cross in the $0^{\circ}6 \times 0^{\circ}.6$ region.
 The panel is smoothed with a Gaussian kernel of 0$^{\circ}$.3.
 Two solid and dashed green circles show the 68\% and 95\% error circles of the best-fit position of
SNR G317.3-0.2, respectively.  The cyan contours are from observations performed using MOST \citep[SNR G317.3-0.2;][]{Whiteoak1996}.}
 \label{Figx}
\end{figure}

In the analysis, we first calculated the test statistic (TS) map, which is a significance map based on the maximum likelihood test statistic, by running {\tt gttmap} centered at the position of SrcX in the 2-500 GeV energy band. 
We found significant $\gamma$-ray radiation with TS value = 32.46 from the direction of SrcX, as shown in the left panel of Figure \ref{Fig1}. 
We chose to add six point sources (P1-P6) with a power-law spectrum to the position of the local maxima of the TS map within $2^{\circ}.2$ to subtract brighter residual radiation in all subsequent analyses. 
We found that $\gamma$-ray radiation was still significant with a TS value of $22.74$, as shown in the middle panel of Figure \ref{Fig1}. 
The fitting parameters of P1-P6 are shown in Table \ref{Table4} (please see the appendix).
After subtracting SrcX, we found that there was no significant residual emission from the location of SrcX, as shown in the right panel of Figure \ref{Fig1}.
Therefore, we suggest that the $\gamma$-ray emission may come from SrcX.
We also checked the 2$^{\circ}$.2$\times$2$^{\circ}$.2 counts map of the ROI using {\tt gtbin}, as shown in Figure \ref{Figx}, with the SrcX position as the center. 
Here, we can see that a certain number of photons are distributed at the position of SrcX. Combined with the significant residual emission from the right TS map in Figure \ref{Fig1}, we suggest that there likely exists a new high-energy $\gamma$-ray source at the location of SrcX.
\begin{table}[!h]
\begin{center}
\caption{Spatial Distribution Analysis for SrcX with Different Spatial Models in the 2-500 GeV energy band}
\begin{tabular}{lccccccc}
  \hline\noalign{\smallskip}
    \hline\noalign{\smallskip}
    Spatial Model & Radius ($\sigma$) & Spectral Index  & Photon Flux & TS
    Value & TS$_{\rm ext}$ & Degrees of Freedom  \\
                  &     degree     &      &    $\rm 10^{-9}  ph$ $\rm cm^{-2} s^{-1}$  &  &     & \\
  \hline\noalign{\smallskip}
   Point source    & ...             & 2.94 $\pm$ 0.51 & 0.39 $\pm$0.10 & 21.68  & - & 4  \\
   2D Gaussian        &  0$^{\circ}$.60  & 2.12 $\pm$ 0.15 & 2.56 $\pm$ 0.36 & 81.25 & 59.57& 5     \\
   
   uniform disk    &  0$^{\circ}$.95 & 2.13 $\pm$ 0.11 & 1.96 $\pm$ 0.33 & 66.56 &44.88 & 5     \\
  \noalign{\smallskip}\hline
\end{tabular}
\end{center}
    \label{Table1}
\end{table}

Here, we checked the $\gamma$-ray spatial distribution of SrcX using uniform disk (2D Gaussian) models with different values of radii ($\sigma$).
Here, the radius ($\sigma$) for the uniform disk (2D Gaussian) model is set in the range of 0$^{\circ}$.1-1$^{\circ}$.2 with a step of 0$^{\circ}$.05.
The fitting results with the highest TS values from the different spatial models are listed in Table \ref{Table1}.
Here, we adopted the Akaike information criterion \citep[AIC;][]{Akaike1974}, which is expressed by the formula $\rm \Delta AIC = AIC_{point} - AIC_{ext}$, to compare the improvement level of spatial models between the point source and extended source.
The AIC values of the point source and extended source are represented using  $\rm AIC_{point}$ and $\rm AIC_{ext}$, respectively. 
A significantly improved AIC value $\approx$ 32.84 was found using the 2D Gaussian spatial template with $\sigma$ = $0^{\circ}.6$, compared to the value obtained via a point-source hypothesis. 
Hereafter, a 2D Gaussian spatial template with a $\sigma$ of $0^{\circ}.6$ was adopted to analyze the $\gamma$-ray emission of SrcX in all subsequent analyses.

\subsection{Variability Analysis} \label{sec:data-results}
\begin{figure*}[!h]
\centering
 \includegraphics[width=\textwidth, angle=0,width=140mm,height=70mm]{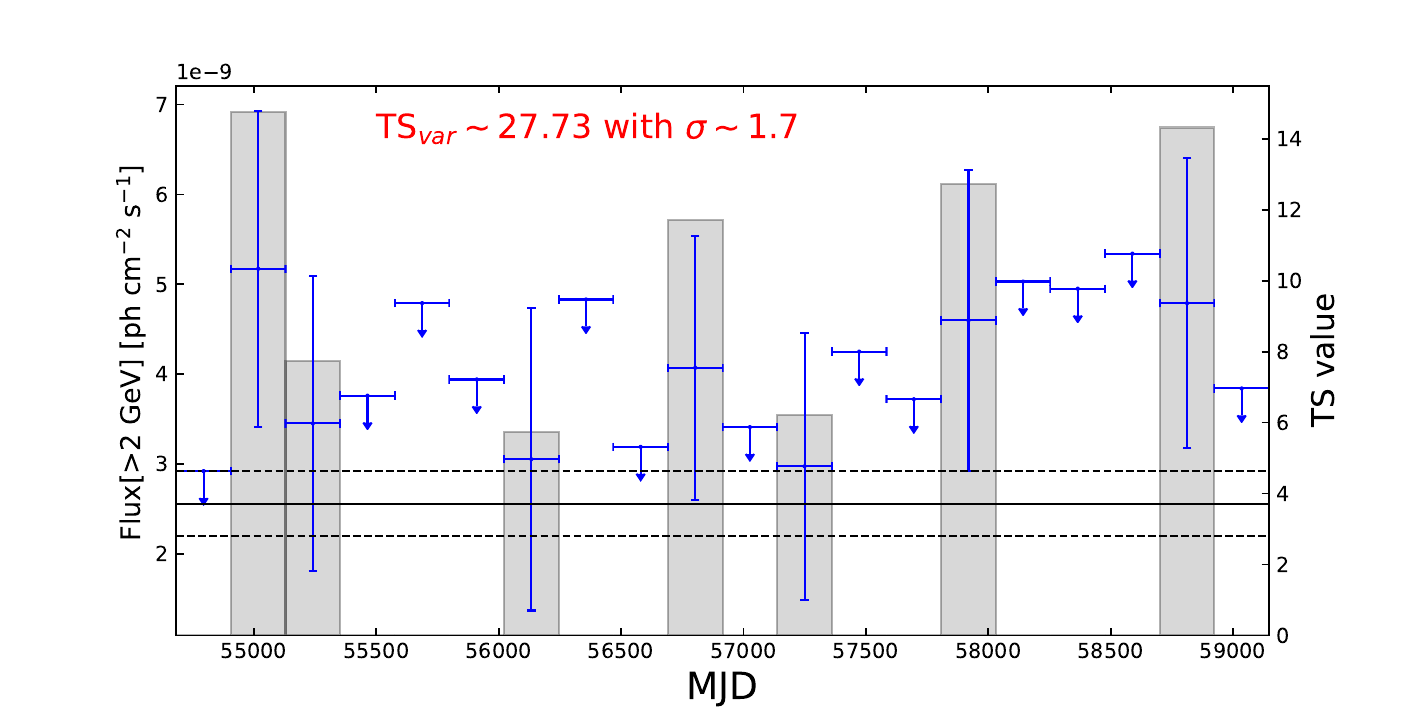} %
 \caption{LC of 20 time bins in the 2-500 GeV energy band for SrcX. The upper limits with the 95\% confidence level are given for the time bin with TS value $<$ 4. The time bins with TS values $>$ 4 are marked by the gray shaded area. The average photon flux with a maximum likelihood of $\sim$ 12.2 years is  represented by the black solid line, and the two black dashed lines indicate its 1$\sigma$ statistic uncertainties.
}
 \label{Fig3}
\end{figure*}

An LC with 20 time bins of $\sim$ 12.2 years in the 2-500 GeV energy band was generated to check the variability of photon flux from SrcX. The variability was estimated  using the variability index TS$_{\rm var}$ from the work by \citet{Nolan2012}. $\rm TS_{var}\geq$ 36.19 was used to identify variable sources at a 99\% confidence level for the LC of 20 time bins.
However, no significant variability from the photon flux was found from SrcX with $\rm TS_{var}=$ 27.73 in the LC, as can be seen from Figure \ref{Fig3}.

\subsection{\rm Spectral Analysis} \label{sec:data-results}
 A source with $\rm TS_{curve}>$ 16 was considered
 significantly curved, where is defined as $\rm TS_{curve}$= 2(log $L$(curved spectrum) - log $L$(powerlaw)) in \citet{Nolan2012}.
  Here, we tested two frequently used models in 4FGL, LogParabola and  PLSuperExpCutoff2\footnote{https://fermi.gsfc.nasa.gov/ssc/data/analysis/scitools/source{\_}models.html}. 
  We found $\rm TS_{curve}\approx$ 0 and $\rm TS_{curve}\approx$ 2 for LogParabola and PLSuperExpCutoff2, respectively.
These results indicate that the spectrum of this source is not significantly curved, which is consistent with those of 45 extended sources with the average value of $\rm TS_{curve}$ =3.2\footnote{Here, we excluded 4FGL J1825.2-1359 with the LogParabola spectral model from this calculation, so its spectrum with $\rm TS_{curve}=21$ is significantly different from other extended $\gamma$-ray sources in FGES.}.  Therefore, we chose a simple PL model as the best spectral model to describe the spectrum of observation data in this analysis.

\begin{figure}[!h]
  \centering\underline{}
   \includegraphics[width=100mm]{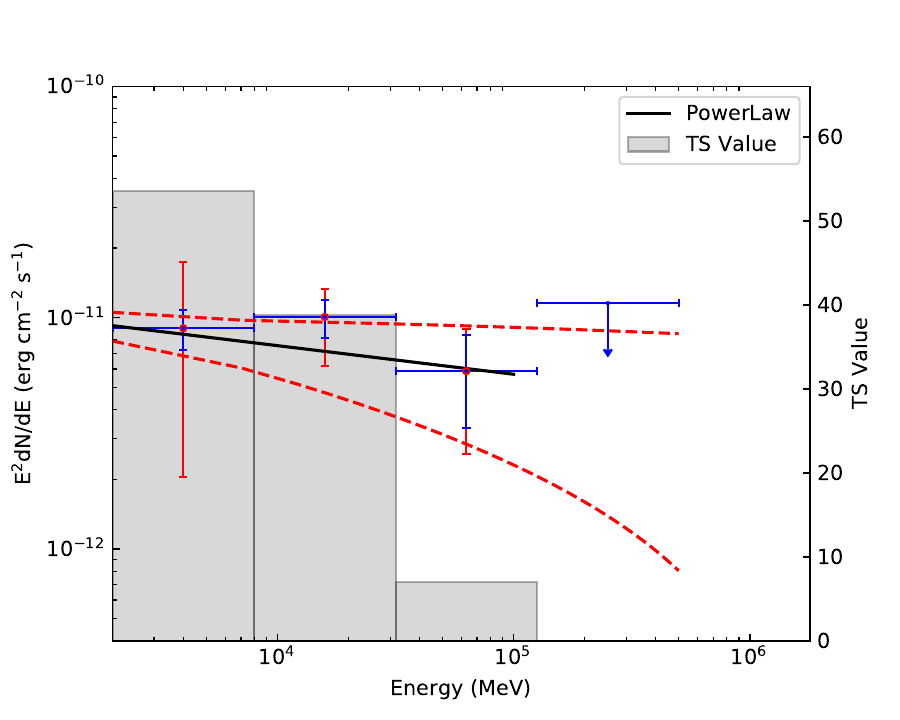}
 \caption{SED of the 2-500 GeV energy band from SrcX. Red points are the result of the Fermi-LAT observation with the total uncertainties including statistical and systematic ones. The systematic uncertainties are from the galactic diffuse emission model and the effective area.
The blue points only consider the statistical uncertainties.
The red dashed and the black solid lines 
represent the 1$\sigma$ statistical uncertainty and the power-law spectrum from the Fermi-LAT global fit, respectively. 
The gray shaded area is used to represent the TS value of the energy bin with the TS value $>$ 4.0. The upper limit of the 95\% confidence level is given for the fourth bin with the TS value $\approx$ 0.}
    \label{Fig4}
\end{figure}

Using the binned likelihood analysis method, the photon flux of the global fit for SrcX was found to be (2.56 $\pm$ 0.36) $\rm\times$ $10^{-9}$ $ \rm  ph$ $\rm cm^{-2} s^{-1}$ with a power-law spectral index of 2.12 $\pm$ 0.15 in the 2-500 GeV energy band.
The 2-500 GeV energy band was selected to generate its spectral energy distribution (SED) with four equal logarithmic bins.
Similar to the method of the global fit, each energy bin was independently fitted  using {\tt gtlike}. An upper limit with a 95\% confidence level was given for the energy bin with a TS value $<$ 4, as shown in Figure \ref{Fig4}. 

Moreover, the systematic uncertainties from the effective area were calculated using the bracketing Aeff
method\footnote{https://fermi.gsfc.nasa.gov/ssc/data/analysis/scitools/Aeff{\_}Systematics.html}. 
The systematic uncertainties of the galactic diffuse emission were estimated by artificially fixing the best-fit value of normalization of the galactic diffuse model from each energy bin to a $\pm$6\% deviation \citep[e.g.,][]{Abdo2009, Abdo2010a,Abdo2010c, Xing2015,Xin2018}. 
The best-fit data are presented in Table \ref{Table2}.
\begin{table}[]
\begin{center}
\caption{The Energy Flux Measurements from SrcX with Fermi-LAT}
\begin{tabular}{lccccc}
  \hline\noalign{\smallskip}
    \hline\noalign{\smallskip}
    E  & Band       & $E^{2}dN(E)/dE$ & TS value \\
 (GeV) &  (GeV)     & ($10^{-12}$erg cm$^{2}s^{-1}$ ) &        \\            
  \hline\noalign{\smallskip}  
    3.99 & 2.0-7.95       &  8.99 $\pm$ 1.77$^{+8.43}_{-6.94}$ &   53.51 \\
    15.86 & 7.95-31.62    &  10.06 $\pm$ 1.90$^{+3.14}_{-3.83}$  &  38.75      \\
    63.06  & 31.62-125.74 &  5.86 $\pm$ 2.54$^{+1.71}_{-2.09}$  &  7.0      \\
    250.74 & 125.74-500.0 &  1.15 &  1.95   \\
  \noalign{\smallskip}\hline        
\end{tabular}
    \label{Table2}
\end{center}
Note. Fluxes with uncertainties from the energy bins with TS value $>$ 4 are given. The first and the second uncertainties from photon energy flux are statistical  and total systematic from the galactic diffuse emission model and the effective area, respectively.  Flux of the fourth energy bin is the 95\% upper limit with TS value $<$ 4.
\end{table}

\section{DISCUSSION and Conclusion} \label{sec:data-results}   

\subsection{\rm Source Detection}

\begin{figure}[!h]
  \centering
  \includegraphics[width=110mm]{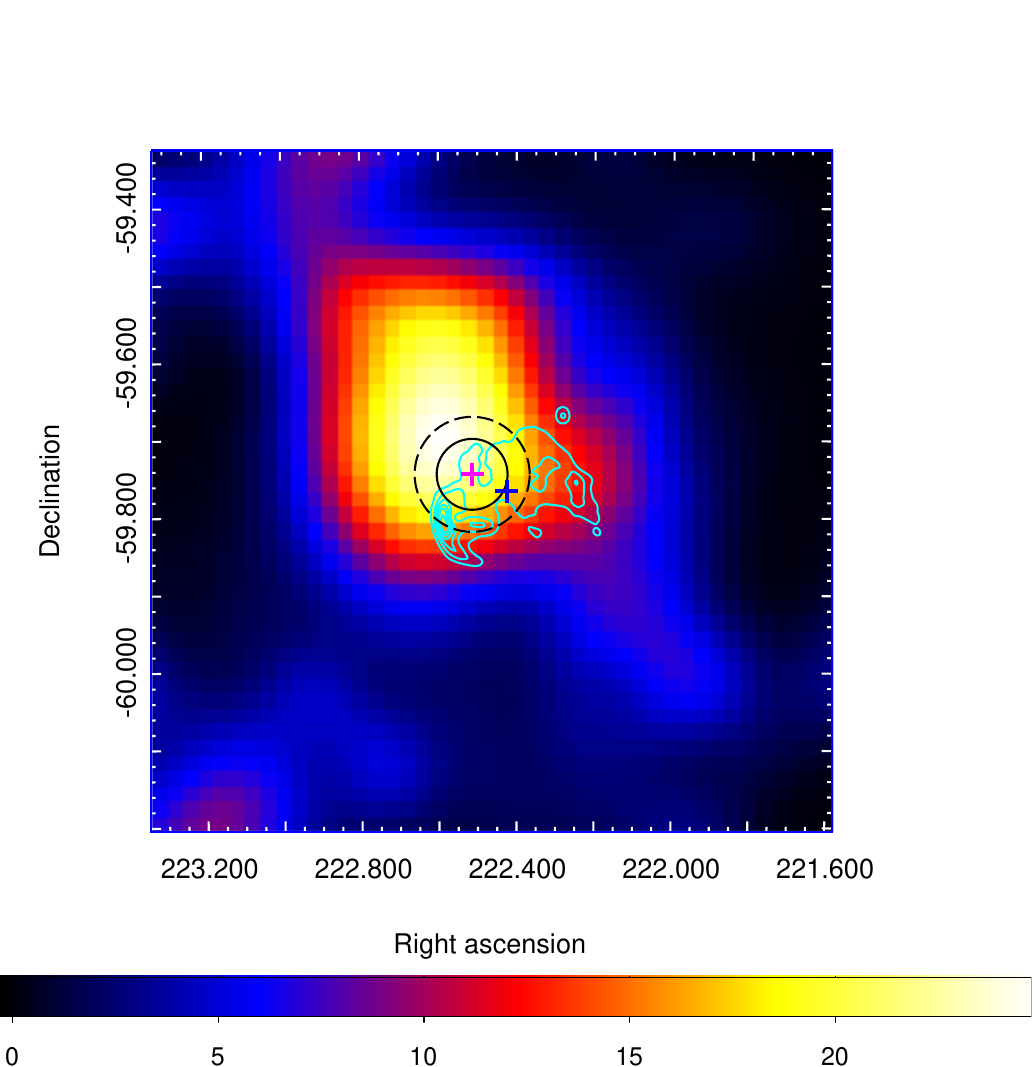}
 \flushleft
\caption{TS map in the 2-500 GeV energy band with 0$^{\circ}$.02 pixel size centered at the best-fit position of SNR G317.3-0.2, where the cyan contours are from the 843-MHz observation of MOST \citep[SNR G317.3-0.2;][]{Whiteoak1996}. 
For the panel smoothed with a Gaussian kernel of 0$^{\circ}$.3, a magenta cross represents the best-fit position of SNR G317.3-0.2 with the spatial template of the point source, and two solid and dashed black circles show the 68\% and 95\% error circles of the best-fit position of SNR G317.3-0.2, respectively. The blue cross represents the original position of SNR G317.3-0.2 from SIMBAD. 
}  
\label{Fig5}
\end{figure}

Through the simulation of the spatial distribution of the $\gamma$-ray source, we found that the 2D Gauss model with index=2.12$\pm$0.15 performs better than a point source template.  For the spectral index range of extended sources, we investigated the Fermi Galactic Extended Source catalog (FGES) from \citet{Ackermann2017} and found that the average value of the spectral index of the extended sources detected above 10 GeV is 2.16. 
We found that the value of spectral index of 2.12$\pm$0.15 corresponding to SrcX is close to this value of 2.16, 
and the power-law spectral characteristic of the new $\gamma$-ray source is consistent with those of the extended $\gamma$-ray sources from FGES in Section 3.2. 
In addition, as shown in the left panel of Figure \ref{Fig5}, 
we find that the location of SNR G317.3-0.2 is within a 2$\sigma$ error radius, and many contours of the 843-MHz SE lobe are within a 2$\sigma$ error radius. 
Meanwhile, SIMBAD\footnote{http://cdsportal.u-strasbg.fr/} and Aladin\footnote{https://aladin.u-strasbg.fr/aladin.gml} were used to investigate whether the region within the 2$\sigma$ error radius of the best-fit position  
corresponds to the currently known GeV $\gamma$-ray sources or candidates such as Pulsar Wind Nebula, Active Galactic Nuclei (AGN), Gamma Burst,  or other SNRs.
 However, we did not find a likely match. 
 Therefore, we suggest that the new $\gamma$-ray emission is likely to originate from SNR G317.3-0.2. 
 
Currently, the most SNRs are considered to be a relatively stable objects in the local Universe. 
They are different from AGNs and do not have short-timescale variability in LC. 
As shown in Figure \ref{Fig3}, in 12.2 years, LC exhibited no significant variability with $\rm TS_{var}$=27.73. 
We then compared the average value of $\rm TS_{var}\approx 10.58$ from 25 SNRs that have been firmly certified in 4FGL and found that the characteristic of no variability of SrcX is consistent with that of most of the SNRs observed in the GeV energy band. This evidence implies that the $\gamma$-ray emission of SrcX is more likely to come from SNR G317.3-0.2, and the new $\gamma$-ray source is probably the GeV counterpart of SNR G317.3-0.2.

\subsection{\rm Model Fitting}

Generally, two scenarios refer to lepton and hadron, respectively, to explain GeV $\gamma$-ray emission production \citep[e.g.,][and references therein]{Condon2017,Xing2019}:

(1) Leptonic scenario, where the $\gamma$-ray emission is induced by the inverse Compton scattering of relativistic electrons.

(2) Hadronic scenario, involving the decay of neutral pion produced by proton-proton interactions.

Through the above discussion, in the case of correlation between SNR G317.3-0.2 and SrcX, for the leptonic scenario, we assume that the population of electrons is given by the formula expressed using a power law with an exponential cutoff (PLEC) as follows \citep[e.g.,][]{Aharonian2006,Xin2019}: 
\begin{equation}
N(E) = N_{\rm 0}\left (\frac{E}{E_{\rm 0}} \right)^{-\alpha}\exp\left( -{\frac{E}{E_{\rm cutoff}}}\right)\;,
\end{equation}
where $N_{\rm 0}$ is the amplitude, $E$ is the particle energy, ${\alpha}$ is the spectral index, $E_{\rm cutoff}$ is the break energy, and $ E_{\rm 0}$ is fixed to 1 TeV.
Here, we consider using a simple one-zone stationary model from {\tt NAIMA} \citep[][and references therein]{Zabalza2015} to explain the Fermi-LAT observations in this work. 
The value of the distance of SNR G317.3-0.2 was adopted to be 9.5 kpc \citep{Guseinov2004}.
We found that the data can be explained well with a reduced $\chi^{2}$ value of 0.01. The best-fit results are presented in Table \ref{Table3}.


For the hadronic scenario, we first investigated the OH maser emission at 1720 MHz around SNR G317.3-0.2 as convincing evidence for verifying the interaction of SNRs with molecular clouds. However, no significant OH maser emission was found by \citet{Frail1996}. 
Moreover, for the case of interactions between the relativistic protons and the surrounding CO gas molecules from the SNR, we investigated the observation results from  the 1.2 meter millimeter-wave (CfA) telescope at the Harvard-Smithsonian center for astrophysics \citep{Dame2001} and found that the molecular cloud in this region is very dense. However, due to a low angular resolution of $\sim$ 8$^{'}.5$ at 115 GHz of the CfA telescope, the interactions from relativistic protons, and CO gas molecules are uncertain thus far. For the uncertain origin of the $\gamma$-ray emission from the region, we also chose to consider the hadronic scenario \citep[e.g.,][]{Condon2017,Xing2019,Zeng2017,Zeng2019}.
Here, we considered the three different gas densities $n_{\rm gas}$ = 1 cm$^{-3}$, 10 cm$^{-3}$, and 100 cm$^{-3}$
 to explain the Fermi-LAT observations, with the population of protons given by formula (1).

\begin{figure}[!h]
  \includegraphics[width=\textwidth, angle=0,width=89mm,height=70mm]{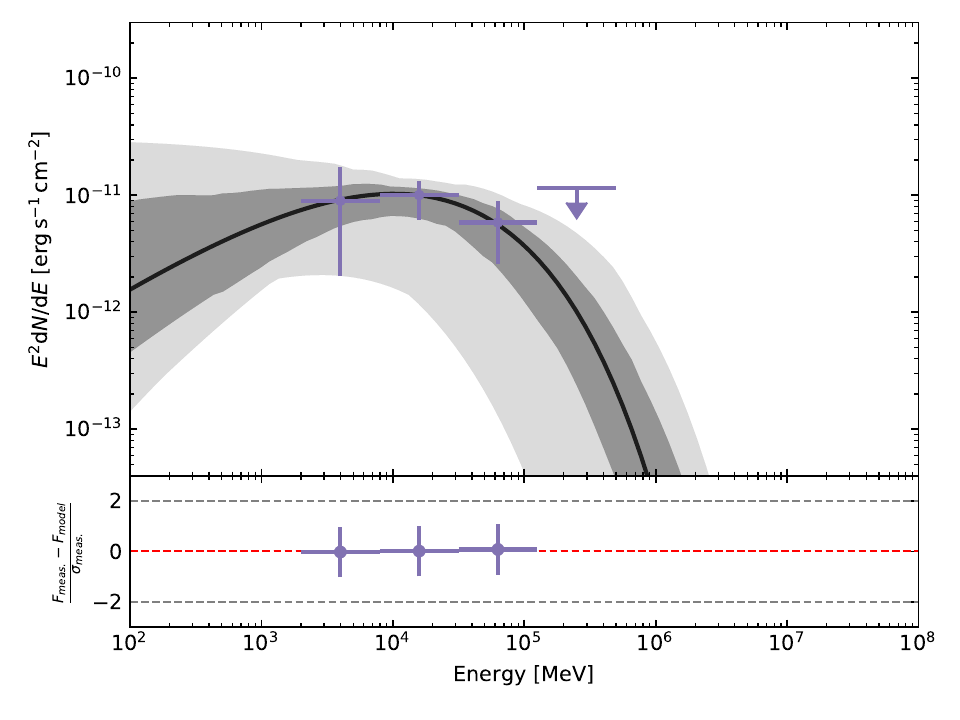}
  \includegraphics[width=\textwidth, angle=0,width=89mm,height=70mm]{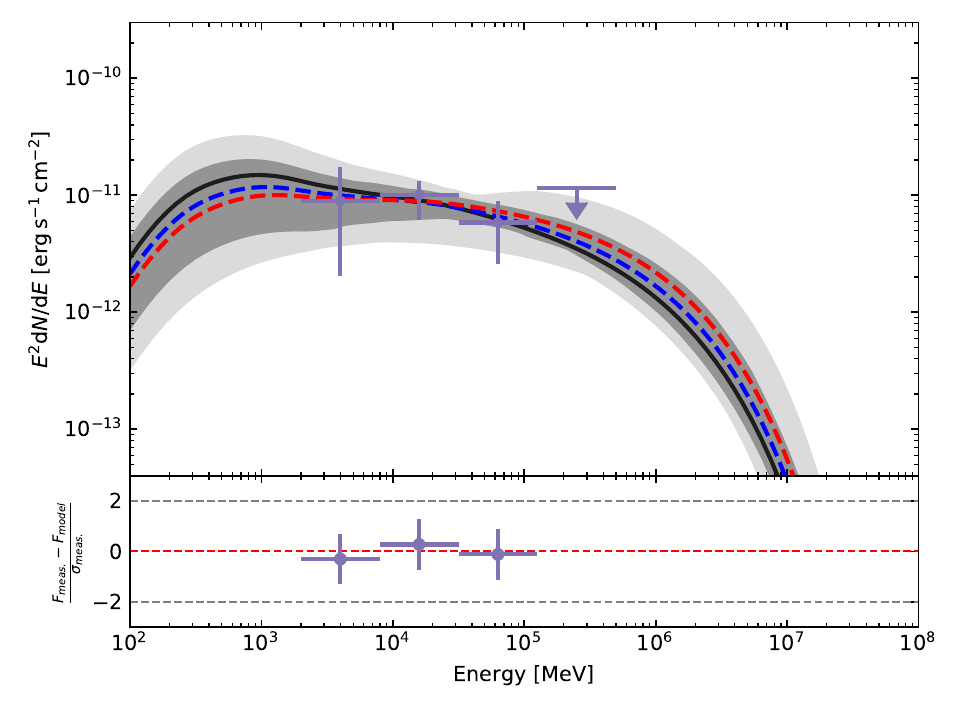}
 \flushleft 
\caption{
The two panels are the best-fit results of the SED of SNR G317.3-0.2 with the leptonic and hadronic models.
The dark and light gray shaded areas show the 1$\sigma$ and 3$\sigma$ confidence regions from the best-fit $\gamma$-ray spectrum marked by a black solid line. 
The lower panel shows the residuals of the measurement with respect to the best fit. 
The left panel shows the leptonic model.
The right panel shows the hadronic model, where the black solid line represents the best-fit result of $n_{\rm gas}$ = 1 cm$^{-3}$,
the blue dashed line represents that of $n_{\rm gas}$ = 10 cm$^{-3}$, and the red dashed line represents that of $n_{\rm gas}$ = 100 cm$^{-3}$. 
}
\label{Fig6}
\end{figure}

Furthermore, {\tt PYTHIA 8}, as a toolkit referring to the cross-section of the proton-proton energy losses and pion production \citep{Kafexhiu2014}, was used to fit the data of SNR G317.3-0.2.
Similar to the leptonic model, we find that the data can be also explained well with the same reduced $\chi^{2}$ value of $\sim$0.2 using the three hadronic models with different gas densities, as shown in the right panel of Figure \ref{Fig6}. 

\subsection{\rm Summary}

\begin{table*}[!h]
\caption{The Best-fit Parameters of Two Models}
\begin{center}
\begin{tabular}{lcccccc}
  \hline\noalign{\smallskip}
    \hline\noalign{\smallskip}
  Model Name       &  $n_{\rm gas}$    & $\alpha$               & $E_{\rm cutoff}$       & $W_{\rm e}$ (or $W_{\rm p}$)  & $\chi^{2}/N_{dof}$   \\
                         &  cm$^{-3}$ &                        & TeV                    & $10^{49}$ erg   &  \\
  \hline\noalign{\smallskip}
  Leptonic model    & --- & $2.07^{+0.56}_{-0.94}$  & $0.96^{+0.42}_{-0.27}$ & $ 3.99^{+0.63}_{-0.66}$   &  $\frac{0.006*2}{4-3}=0.01$   \\
      \noalign{\smallskip}\hline 

  Hadronic model    & 1 & $2.14^{+0.21}_{-0.24}$  & $10.1^{+1.33}_{-1.49}$ & $ 394.15^{+76.79}_{-67.12}$   &    $\frac{0.1*2}{4-3}=0.20$ \\
  
    \hline\noalign{\smallskip}
 Hadronic model    &  10                   &$2.26^{+0.24}_{-0.31}$  & $9.77^{+1.62}_{-1.29}$ & $28.43^{+8.04}_{-6.38}$  & $\frac{0.09*2}{4-3}=0.18$   \\
     \hline\noalign{\smallskip}
 Hadronic model    &  100                   &$2.14^{+0.27}_{-0.35}$  & $9.93^{+1.58}_{-1.50}$ & $3.57^{+0.34}_{-0.66}$  & $\frac{0.1*2}{4-3}=0.20$   \\
  \noalign{\smallskip}\hline  
\end{tabular}
\end{center}
\label{Table3}
Note: the upper limit of the fourth bin is taken into account in the fit \citep[e.g.,][]{Abdalla2018}. For the values of $W_{\rm e}$ (or $W_{\rm p}$) with  electronic (or protonic) kinetic energy of above 2 GeV, first, we fixed the best fitting values of the rest of the parameters and finally independently fitted $W_{\rm e}$ (or $W_{\rm P}$).
\end{table*}

We analyzed the $\gamma$-ray radiation from the region of SNR G317.3-0.2 by using the latest Fermi-LAT Pass 8 data and found that its photon flux was $\rm (2.56\pm0.36)\times 10^{-9}  ph$ $\rm cm^{-2} s^{-1}$ with $\sim$ 8.13$\sigma$ in the 2-500 GeV energy band.
Further, we found that its spectral characteristic, which has a spectral index of 2.12$\pm$0.15 with the power-law spectrum, is in good agreement with those of other extended sources.
Meanwhile, the 2D Gaussian $\gamma$-ray spatial distribution showed  better improvement than other spatial models by AIC.
Its spatial position coincided well with that of the 843-MHz radio band, and no other known GeV $\gamma$-ray sources or candidates were found in the region of SNR G317.3-0.2.
We also analyzed the variability of LC in $\sim$ 12.2 years and found that there was no significant variability of photon flux, which implies the characteristic of no variability of its LC is in good agreement with the SNRs observed thus far. Therefore, we suggest that the new $\gamma$-ray source is probably the GeV counterpart of SNR G317.3-0.2.
Moreover, the leptonic and hadronic models with the same PLEC particle population could well explain the GeV radiation spectrum in the 2-500 GeV energy band, independently.

\section{Acknowledgments} 
We sincerely appreciate the support for this work from
the National Key R\&D Program of China under Grant No. 2018YFA0404204, the National Natural Science Foundation of China (NSFC U1931113, U1738211, 11303012) ,the Foundations of Yunnan Province (2018IC059, 2018FY001(-003), 2018FB011), the Scientific Research Fund of Yunnan Education Department (2020Y0039).

\section{APPENDIX}
Table 4 contains relative fitting information from the six point sources added in Section 3.

\begin{table}[!h]
\caption{The Best-fit Parameters of the $\gamma$-ray Excesses of P1-P6 in the 2-500 GeV energy band}
\begin{center}
\begin{tabular}{cccccc}
  \hline\noalign{\smallskip}
    \hline\noalign{\smallskip}
      Source  Name        & $\rm TS$    &  $R.A.$  &  $Decl.$  & $N_{0}$ & $\Gamma$   \\
                    &         &   [deg]  & [deg]     & $10^{-13}$&   \\
  \hline\noalign{\smallskip}
         P1    &    18.62  & 224.19 & -59.50 & 4.47 $\pm$ 1.48 &   3.33 $\pm$ 0.49\\
         P2    &    23.01  & 221.52 & -59.31 & 2.55 $\pm$ 1.02  & 2.62 $\pm$ 0.32\\     
         P3    &    20.71  & 221.20 & -60.03  & 4.45 $\pm$ 1.51  & 3.12 $\pm$ 0.42\\ 
         P4    &    19.27  & 220.94 & -60.42  & 1.62 $\pm$ 0.70 & 2.23 $\pm$ 0.23\\    
         P5    &    9.99   & 221.78 & -59.59  & 0.40 $\pm$ 0.35 & 1.78 $\pm$  0.30 \\ 
         P6    &    13.13  & 223.05 & -59.86  & 0.48 $\pm$ 0.35  & 1.74 $\pm$ 0.27\\ 
  \noalign{\smallskip}\hline
\end{tabular}
\end{center}
\textbf{Note}:
The fitting results of P1-P6 with a power-law spectrum, $dN/dE=N_{0}(E/E_{0})^{-\Gamma}$, here $E_{\rm 0}$ is fixed to 2 GeV.
\label{Table4}
\end{table}

\end{document}